\documentclass[unnumsec,webpdf,modern,large]{oup-authoring-template}

\onecolumn 

\graphicspath{{Fig/}}

\theoremstyle{thmstyleone}%
\theoremstyle{thmstyletwo}%
\theoremstyle{thmstylethree}%

\begin{document}

\journaltitle{This preprint has been submitted to Journal of Cybersecurity}
\DOI{DOI HERE}
\copyrightyear{2022}
\pubyear{2023}
\access{Advance Access Publication Date: Day Month Year}
\appnotes{Paper}

\firstpage{1}


\title[Human Cyber-Defense Behavior]{ Learning About Simulated Adversaries from Human Defenders using Interactive Cyber-Defense Games}

\author[1]{Baptiste Prebot}
\author[1]{Yinuo Du}
\author[1,$\ast$]{Cleotilde Gonzalez\ORCID{0000-0002-6244-2918}}

\authormark{Baptiste Prebot et al.}

\address[1]{\orgdiv{Social and Decision Sciences Department}, \orgname{Carnegie Mellon University}, \orgaddress{\street{Pittsburgh}, \postcode{15213}, \state{PA}, \country{USA}}}

\corresp[$\ast$]{Corresponding author. \href{email:coty@cmu.edu}{coty@cmu.edu}}

\received{Date}{0}{Year}
\revised{Date}{0}{Year}
\accepted{Date}{0}{Year}



\abstract{Given the increase in cybercrime, cybersecurity analysts (i.e. Defenders) are in high demand. Defenders must monitor an organization's network to evaluate threats and potential breaches into the network. Adversary simulation is commonly used to test defenders' performance against known threats to organizations. However, it is unclear how effective this training process is in preparing defenders for this highly demanding job. In this paper, we demonstrate how to use adversarial algorithms to investigate defenders' learning of defense strategies, using interactive cyber defense games. Our Interactive Defense Game (IDG) represents a cyber defense scenario that requires constant monitoring of incoming network alerts and allows a defender to analyze, remove, and restore services based on the events observed in a network. The participants in our study faced one of two types of simulated adversaries. A Beeline adversary is a fast, targeted, and informed attacker; and a Meander adversary is a slow attacker that wanders the network until it finds the right target to exploit. Our results suggest that although human defenders have more difficulty to stop the Beeline adversary initially, they were able to learn to stop this adversary by taking advantage of their attack strategy. Participants who played against the Beeline adversary learned to anticipate the adversary and take more proactive actions, while decreasing their reactive actions. These findings have implications for understanding how to help cybersecurity analysts speed up their training.
 }
\keywords{Cyber defense, human behavior, cyber adversary, interactive games}


\maketitle

\section{Introduction}
The rapidly evolving attack capabilities to deploy increasingly sophisticated cyber attacks of unprecedented speed and scale require well-trained cybersecurity experts (i.e., defenders, analysts) to address evolving sophisticated attack strategies [\citenum{li2021comprehensive,thanh2019survey}]. Cyber analysts are responsible for protecting an organization's computer network and digital assets. The job of these defenders consists of a wide variety of network-dependent tasks, including the examination of a large number of alerts to identify intrusion activities and determine whether a network is under attack, the detection of flaws in the organization's security, the development of appropriate protections, and, of course, the mitigation of threats. These activities often include making time-sensitive decisions that may involve disrupting the organization's work in order to protect their information.

To investigate these cyber situations, evaluate defense algorithms and strategies, and train defenders against new threats, cyber wargaming and adversary simulation are common practices \citenum{colbert2020game, ferguson2018tularosa}. Wargaming exercises mimic a potential threat to an organization by using threat intelligence to define what actions and behaviors an adversary may use. Wargaming emulators build scenarios that capture certain aspects of tactics, techniques, and procedures, to help test the efficacy of defense and identify vulnerability of the network \citep{applebaum2016intelligent}. Human defenders are usually recruited to interact with adversarial simulated scenarios to help them learn from such an interaction \citep{kavak2021simulation,varshney2011live}.

Despite a growing interest in cyber defense behaviors in recent years \citep{gutzwiller2016task,veksler2020cognitive,veksler2018simulations,cranford2021towards,johnson2021decision}, our understanding of the cognitive demands faced by cyber analysts is still limited \citep{gonzalez2014cognition}.
Many factors in adversarial behavior may influence defense strategies. For example, aggressor personality traits are known to influence their cyber attack behaviors \citep{Triad_JONES2021106799, Triad_CURTIS2018174}: Long-term mimicry deception and Machiavellianism were found to be predictors of stealthy attacks, while narcissism and psychopathy were associated with shorter and more aggressive attacks (i.e., “brute force”). 

Human-in-the-loop cyber defense laboratory research is required to study both defensive and offensive cyber operations and to develop training protocols tailored to different types of attack strategies \citep{gutzwiller2015human}. However, conducting meaningful laboratory research with simulated adversaries to study defender behavior is challenging. 
Participants with the skills and knowledge required to test highly technical tasks and sophisticated adversaries are hard to find and are often too busy to provide their time to test simulated adversaries \citep{veksler2020cognitive,BUCHLER2018}.
The design of simulated adversaries with high-fidelity in terms of techniques also requires extensive threat intelligence collected through long-term tracking and clustering of intrusion activities \citep{strom2018mitre}. 
Given the continuous evolution of network environments and potential adversaries, it is also unrealistic to derive a future-proof defense strategy at the granularity of concrete techniques.

To help mitigate this challenge, researchers have been using simulation tools and simplified games \citep{gonzalez2005use} to study the offensive and defensive sides of cyber deception \citep{aggarwal2020hackit,cranford2021towards}, to understand how the general public classifies phishing emails \citep{Triad_CURTIS2018174,singh2019training}, to investigate how the cyber security knowledge of the attacker affects the identification of attacks \citep{ben2015effects}, and to study the behavior of the attacker under different levels of uncertainty about the attacker's strategy \citep{moisan2017security}.
In this work, we adopt the \textit{Intrusion kill chain} model \citep{hutchins2011intelligence} to simplify sophisticated cyber attacks into three tactical phases \textit{Establish initial foothold, Propagate through network}, and \textit{Act on objectives} \citep{zhang2021three}. Consequently, countermeasures such as \textit{Monitor, Analyze, Remove, Restore} are adopted to disrupt each phase of the attack lifecycle.
By pairing defenders with various adversarial strategies constructed with the above tactics, we can learn about the behaviors of human defenders and their processes to address different types of attackers and adapt to dynamic network environments.


However, there is a lack of research on investigating the impact of different adversarial strategies on defense behaviors and the development of defense strategies.
Most adversarial cybersecurity games rely on game-theoretic approaches to determine the best defense strategies. These methods often only consider a particular adversary and assume that opponents act "rationally" (i.e., exhibit optimization behavior). These techniques assume the availability of information to adversaries rather than uncertainty, as is more common in real life, and provide individuals with an exact payoff matrix \citep{tambe2011security,abbasi2016know}. This leads to a misrepresentation of the reality of the highly dynamic cyber environment, where analysts must work with incomplete and flawed information.
While game-theoretic approaches can be useful in determining the optimal defense strategies against known attacks, they provide an unrealistic representation of the attacker's intentions \citep{aggarwal2015cyber,nochenson2012simulation,do2017game}; leading to instantiation that might ultimately perform poorly in dynamic cyber defense environments against unfamiliar adversaries \citep{do2017game,attiah2018game,wang2016survey}. 

\subsection*{Goals and Research Method}
In this research, we address the question of how human defenders behave against different attack strategies and how it affects the emergence of defense strategies. We defined two adversarial strategies in a particular but generic network setting. One adversarial strategy (i.e., Meander) was stealthy; and another one was direct and speedy (i.e., Beeline), reflecting two attack personality or goals. 

In a recent experiment, \cite{hfesIBL} confronted an instance-based learning model, a form of cognitive model that is designed to mimic human decisions \citep{Gonzlez2003InstancebasedLI}, with both of these adversarial strategies. The simulation experiment captured the differences in attack strategies and their effect on defenders outcomes. Mainly, the Beeline strategy resulted in the worst performance for the model than the Meander strategy. However, human data was not available to validate these observations. 

We designed an Interactive Defense Game (IDG) in a cybersecurity scenario and conducted a laboratory study to test human defense behavior against the two adversarial strategies. 
Similarly to \cite{hfesIBL}, we expect participants who face a Beeline strategy to have more difficulty defending their network against intrusions than participants who face the Meander strategy. 


\section{Interactive Defense Game}\label{sec2}

The Interactive Defense Game (IDG) is a web-based interactive cyber defense game developed to study how human defenders make decisions in a cybersecurity situation. The IDG does not require any installation and can be played remotely using a web browser. \footnote{Demo of the game: \url{http://janus.hss.cmu.edu:8084/}}
It provides human participants with a graphical interface to observe network events and analyze the information about a computer network, similar to the way Intrusion Detection Systems (IDS) present network events to human defenders. IDS are common tools to monitor the activities on a network and to help detect possible intrusions or attacks \cite{gonzalez2014cognition}. 

\subsection{The task of a cyber defender in the IDG}
In the IDG, participants play the role of cybersecurity analysts hired by a fictitious manufacturing company to protect their computer network from external malicious activity. The network we use is a simplified version of common corporate network topologies. It is composed of hosts, staff computers, and servers grouped in subnets. Attackers are trying to gain access to the Operational Server (Op\_Server0) to steal information and disrupt production. The easiest way for them to do so is to enter the network through one of the staff computers on the first subnet and progressively make their way up to the critical Op\_Server0 by gaining administrator access to every host on their way.

Each host on which an attacker got administrator-level access costs the defenders some points. The goal of the defender is to minimize the number of points lost.

\begin{figure}[!t]
    \centering
    \includegraphics[scale=.4]{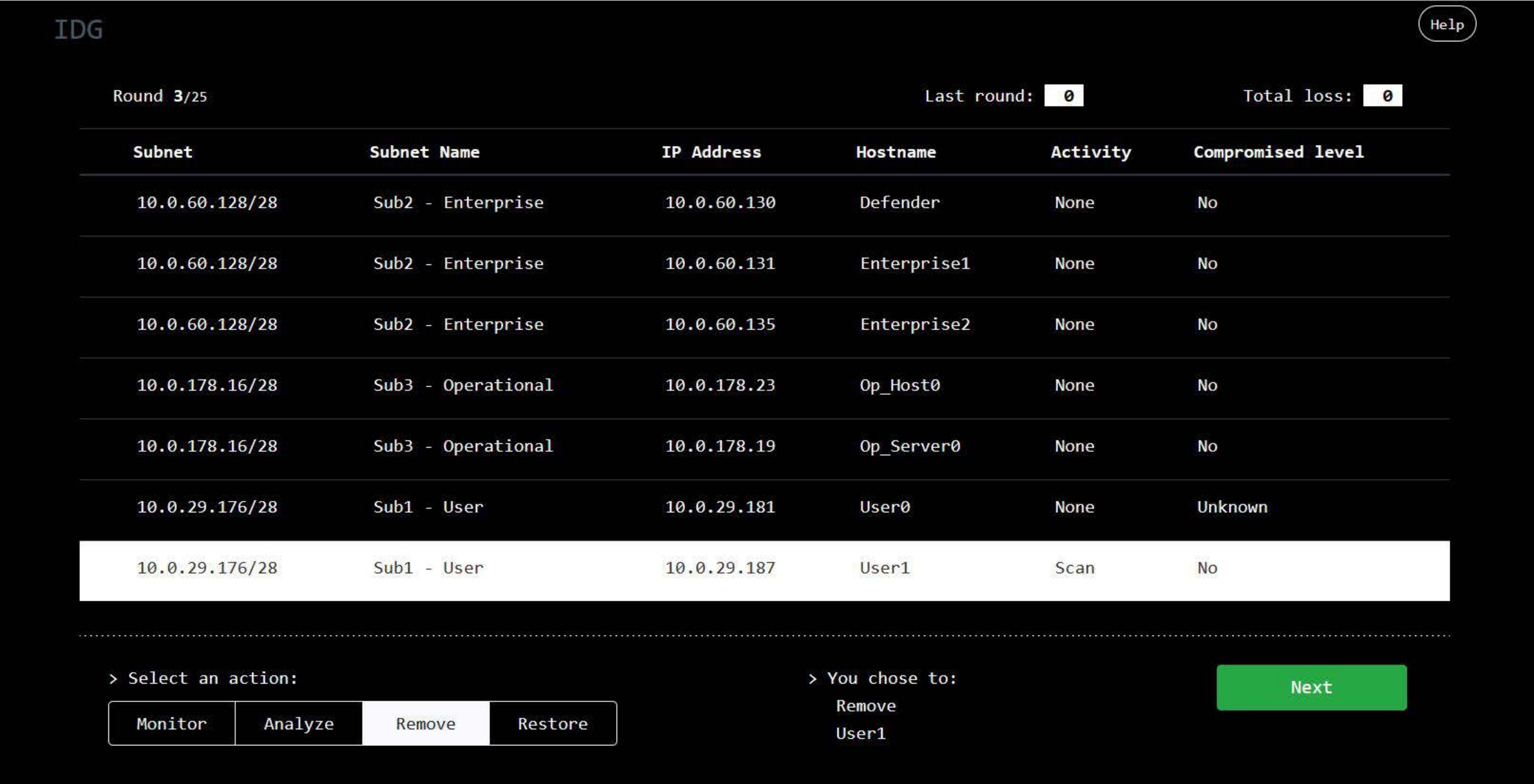}
    \caption{Illustration of the Interactive Defense Game user interface.}
    \label{fig:UI}
\end{figure}

To perform this task, the defenders use the IDG interface shown in Fig. \ref{fig:UI}. They must actively monitor the activity of the network to try to identify malicious activity and take actions to block the progression of the attacker. The hosts of the network are characterized by the subnet to which they belong, an ip address, and a host name. Additionally, the system provides the defenders with two dynamic piece of information about each host, the Compromise level and the Activity. 
When targeting a host, the attacker will first try to gain user-level access to the machine, then try a privilege escalation to gain administrator-level access, and progress to the next target in the network. The \textit{Compromise level} indicates the status of infection of the host.
The second dynamic element provides information about the last \textit{Activity} detected by the system, like scans or exploitation attempts performed by the attacker on this host. However, not all attacker's activities can be detected by the system. More advanced actions, for example privilege escalation attempts and their consequences, are automatically detected.
Thus, the defenders have to understand the observable activity and compromised levels to anticipate future actions of the attackers.

Based on these observable elements defenders can select among a set of actions represented in buttons on the bottom right of the screen: Monitor, Analyze, Remove and Restore. Human defenders can select a host by clicking on its row in the table and then choose one of the four actions to perform on that particular host. Only the Monitor action does not require to select a target, it applies to the whole network.

Then, after clicking on the 'Next' button, the selected action takes effect, and the defender can see the result (i.e. amount of points lost) from the execution of that action in the 'Last round' value. A new and updated version of the environment is presented to the human defender, demonstrating the new state (activity and compromised levels) of the network elements. The 'last round' outcome provides immediate feedback regarding the effectiveness of the past action, and the 'total loss' presents the human defender with a cumulative account of the loss during the game.
Each game lasts a fixed number of \textit{steps}, each step representing one action. 

\subsection{Defense Scenario and Attack Strategies}
\begin{figure}[!t]
    \centering
    \includegraphics[scale=.4]{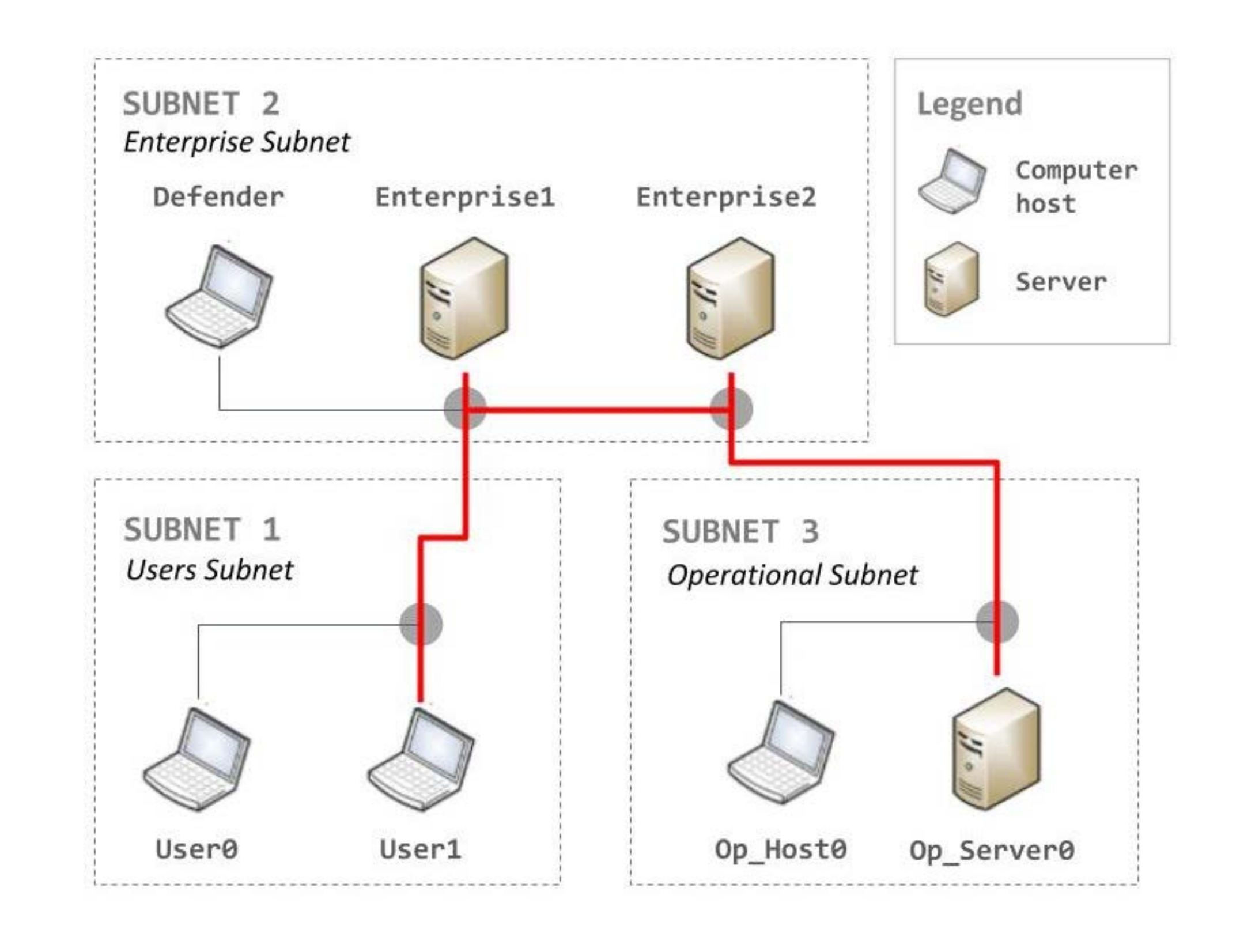}
    \caption{Topology of the network being defended in the IDG scenario. The red line represents the path any attacker needs to take to access the Operational Server.}
    \label{fig:network}
\end{figure}

Human defenders in the IDG are asked to defend a computer network against a red agent. The specific network we used in this scenario is illustrated in Fig. \ref{fig:network}.

The network is composed of 7 hosts (4 computer hosts and 3 servers) distributed across 3 subnets. Subnet 1 consists of user hosts that are not critical, subnet 2 consists of enterprise servers designed to support the user activities on subnet 1, and subnet 3 contains the critical operational server and an operational host.

Two types of attack strategy are implemented. They differ by the assumption of the attacker's prior knowledge and illustrate attack behaviors that may result from differences in the attacker's personality traits \cite{Triad_JONES2021106799, Triad_CURTIS2018174}. In the \textit{Beeline} strategy, attackers route directly through subnet nodes to the Operational Server. The \textit{Meander} strategy does not assume any prior knowledge of the network from the attacker. Attackers following this strategy wonder through the network, trying to gain privileged access to every host in a subnet before advancing further into the network.
As a consequence, the Beeline strategy is a direct, rapid, and targeted strategy that can reach the Operational server faster than an attacker following the Meander strategy. For the defender, the implications are a higher theoretical maximum loss against Beeline (-160) than against Meander (-100). These are the results of a completely passive defender. The Beeline strategy can also result in more disruptions and perhaps longer recovery times if the defender has more difficulty detecting such disruptions.

\section{Methods} \label{sec:experiment}

\subsection{Experimental Design}
The goal of this experiment is to compare the behavior of human defenders faced with the two types of attack strategy discussed above: \textit{Beeline} and \textit{Meander}.

Given the characteristics of the Beeline strategy that can be faster and more damaging to defenders compared to the Meander strategy, we expected that defenders would initially perform worse against Beeline than against Meander. This hypothesis was preregistered with the Open Science Framework \footnote{\url{https://osf.io/u3nfh}}.

\subsection{Participants}
Participants were recruited through Amazon Mechanical Turk to participate in a cybersecurity study. The study was advertised to last between 35 and 45 minutes. The time it took across participants was M = 47.02$\pm$13.16 Minutes. Participants received a base compensation of \$4.5, and up to \$5.6 in bonus payment (M = 3.96$\pm$1.39) based on their final score\footnote{As the score used in this experiment is negative (loss),the bonus payment was calculated by using the difference to the maximum possible loss and attributing 0.005\$ per point: bonus=(\textit{total loss}+1120)*0.005.}. 

120 participants (89 male, 30 female, 1 N/A) aged 21-65 years-old (M = 36.77 $\pm$ 11.00) completed the study.
12 of the 120 participants (10\%) had more than 5 years of experience in the network operation and security area and at least a Master's degree in a related field. 

Each participant was randomly assigned to face one of the two adversarial strategies. 

\subsection{Procedure}
After giving their informed consent and completing a demographic questionnaire, participants received instructions for the task followed by a short quiz to verify their basic understanding of the task instructions. Participants had to correctly answer all the questions before moving on to the next step of the experiment. Participants received feedback on the accuracy of their responses and were allowed to modify their responses if they were incorrect. There was no limit in the number of attempts the participants had to answer the questions correctly.  However, we recorded the score of their first attempt and the number of times they tried to answer the questions.

Next, participants watched a video introduction to the IDG, explaining the interface, the game controls, and the dynamics of an episode.  

Then, participants performed the task consisting of two phases: (1) a practice session and (2) a main task.
The practice session consisted of two short episodes (i.e. games) of 10 steps each. The practice episodes were intended to familiarize participants with the interface and game controls. Each of the practice episodes was associated with one of the attacker strategies; however, since the two attack strategies do not differ significantly during the first 10 steps, the participants did not have enough information to discriminate between the two adversarial strategies during the practice session.

Following the practice session, the participants performed the main task consisting of 7 episodes of 25 steps each. No time restrictions were imposed. The experimental conditions were kept constant throughout the episodes, which means that each participant played 7 episodes against the same adversarial strategy. The initial state of the network was the same for all participants and for each of the episodes.

Subsequently, participants completed a post-experiment survey composed of two parts: (1) feedback on their performance and perceived strategy and (2) their experience in computer science and cyber defense. Finally, the participants received their final score and were dismissed. The experimental instructions, quiz, and surveys, along with the data and analysis scripts, can be accessed at \url{https://osf.io/u3nfh}. 

\subsection{Outcome and Process Metrics} 
We measured the outcome of the defense performance in the IDG using three metrics:
\begin{itemize}
    \item \textbf{Loss}: total number of points lost by the defender during the scenario. For reference, the maximum loss per episode resulting from Beeline actions is -160, while it is -100 against Meander. 
    \item \textbf{Disruptions}: number of server disruptions that occur within each episode. One disruption represents a set of consecutive steps between a successful impact attack on the operating server and the successful recovery by the defender. 
    \item \textbf{Recovery Time}: the average number of steps per episode that the defender takes to remove the attacker from the operational server after it is disrupted.
\end{itemize}

We also measured defense process behaviors in addition to defender decisions (i.e, which action is chosen in each step). The attacker actions were also logged for each step and were used to analyse the human behaviors and strategies of defense:

\begin{itemize}
    \item \textbf{Proportion of Defense actions}: number of times that each of the four defense actions - \textit{Analyze, Monitor, Remove, Restore}- is used by a participant within each episode, divided by the length of the episode (25 steps).
    \item \textbf{Proportion of Attacker's targets}: number of times each host or subnet is being targeted by the attacker within each episode, divided by the length of the episode (25 steps). This is indicative of the attacker's path in the network.
    \item \textbf{Proportion of Defense strategy}: the frequency with which each of three coded strategies of defense have been used (\textit{Reactive, Proactive, Passive}) within each episode. Details of calculations of these strategies are presented in section \ref{DefenseStratSection} below.
\end{itemize}


\section{Results}
 
\subsection{Outcome Metrics}
Table \ref{tab:descriptive_stat} presents the average loss, the number of disruptions, and the recovery time of the participants who played against the Beeline attack strategy and those who faced the Meander attack strategy.


     \begin{table}[!t]
        \caption{Descriptive statistics (mean $\pm$ standard deviation) regarding average loss, number of disruptions,recovery time and success rate per episode. For contextualization, the maximum loss per episode is -160 against Beeline, -100 against Meander.}\label{tab:descriptive_stat}
        \begin{tabular*}{\columnwidth}{@{\extracolsep\fill}llll@{\extracolsep\fill}}
        \toprule
         & Beeline & Meander\\
        \midrule
        Loss  & -56.12 $\pm$ 50.73 & -34.76 $\pm$ 30.40 \\
        Disruptions & 0.94 $\pm$ 0.81 & 0.49 $\pm$ 0.52 \\
        Recovery Time (steps) & 2.75 $\pm$ 3.55 & 1.31 $\pm$ 1.69 \\ 
        \botrule
        \end{tabular*}
    \end{table}
    
These observations corroborate some expected differences between the two attack strategies in each of the three metrics for outcome performance. In general, the participants lost more points against the Beeline strategy than against the Meander strategy. The average number of disruptions to the operational server within one episode was larger when playing against the Beeline than when playing against the Meander strategy. It also took more steps within an episode to remove the attacker from the operational server when disrupted by the Beeline than the Meander attacker. 

\begin{figure}[!t]
    \centering
    \includegraphics[scale=.9]{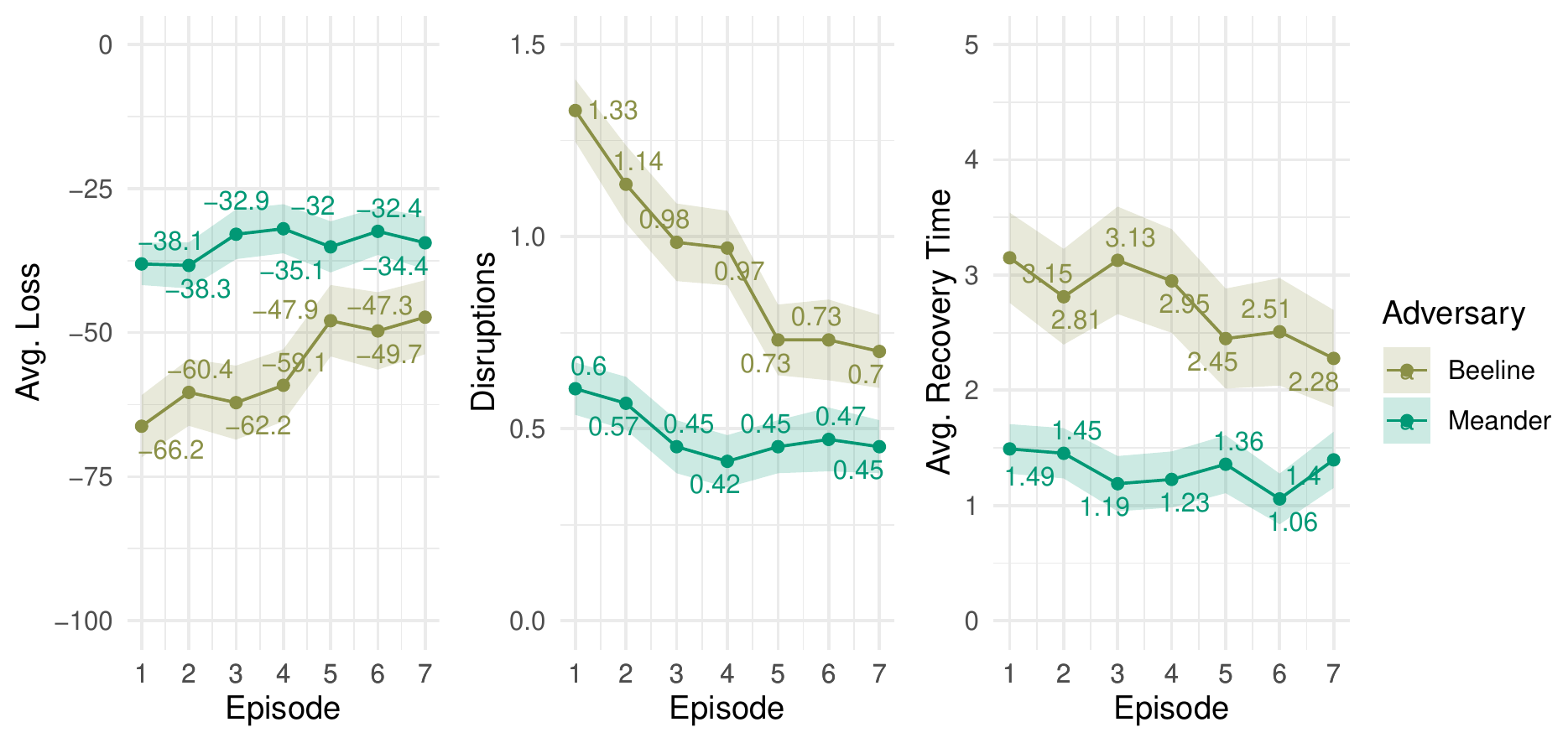}
    \caption{Outcome metrics over time with standard error of the mean. From left to right: Loss; Disruptions; Recovery time.}
    \label{fig:Loss}
\end{figure}

We analyzed the outcome metrics over episodes to determine whether the defenders improve with practice against each of the two adversaries. Fig. \ref{fig:Loss} shows the average of each of the three outcome metrics per episode. Generally, we observe more stability over episodes in the participants' outcomes against the Meander adversary than against the Beeline adversary. In other words, the initially poorer performance of participants against a Beeline adversary improves with more practice with this adversary, while the performance of participants against the Meander adversary does not improve much over episodes.

The participants' \textbf{losses} are lower and relatively more stable against the Meander adversary; however, the participants' losses are larger against the Beeline adversary, and they decrease with more practice against this adversary. In addition, the average number of server \textbf{disruptions} is initially higher for participants confronted with the Beeline adversary compared to those confronted with the Meander adversary. However, the number of disruptions decreases with more episodes against the Beeline adversary. A similar result is observed in the \textbf{average recovery time} per episode; where the time is longer for participants playing against the Beeline adversary compared to the Meander adversary, but it decreases with more episodes.


These observations were tested using mixed-effects analysis of variance (ANOVAs) that included the adversary as a between-subjects factor, the episode as a within-subjects factor, and their interaction. The results for each of the three outcome metrics are reported in Table \ref{tab:anova_outcome}.

\begin{table}[!t]
        \caption{Results of the mixed ANOVAs regarding the effect of adversary type and episodes on outcome metrics}\label{tab:anova_outcome}
        \begin{tabular*}{\columnwidth}{@{\extracolsep\fill}lllllllll@{\extracolsep\fill}}
        \toprule
         Metric & & NumDF & DenDF & F value & p & p.signif & $\eta^2$\\
        \midrule
        Loss & & & & & & &\\
        & Adversary  &  1.00 & 117.00 & 8.44 & .004 & ** & .06\\
        & Episode  &  4.45 & 520.94 & 5.99 & $<.001$ & *** & .01\\
        & Adversary:Episode & 4.45 & 520.94 & 3.54 & .005 & ** & .01\\
        Disruptions & & & & & & &\\
        & Adversary  & 1.0 & 117.00 & 24.24 & $<.001$ & *** & .10\\
        & Episode  & 5.1 & 596.38 & 10.08 & $<.001$ & *** & .04\\
        & Adversary:Episode & 5.10 & 596.38 & 4.34 &$<.001$ & *** & .02\\
        Recovery time & & & & & & &\\
        & Adversary  & 1.0 & 117.00 & 8.87 & .004 & ** & .06\\
        & Episode  & 4.78 & 559.48 & 2.09 & .068 &  & .00\\
        & Adversary:Episode & 4.78 & 559.48 & 1.62 & .157 &  & .00\\
        \botrule
        * p $< 0.05$, ** p $< 0.01$, *** p $< 0.001$.
        \end{tabular*}
\end{table}

Statistical results indicate that the loss, disruptions, and recovery time of the defenders are significantly different when facing the Beeline or Meander adversaries. With the exception of \textbf{average recovery time}, we also found consistent significant effects of the episode and the interactions between the adversary and the episode in the Loss and Disruptions.  

Post-hoc 1-way ANOVAs for each of the metrics confirm what we observed in the figure: loss and disruptions improved over the course of episodes \textit{only} when participants confront the Beeline adversary, but not when paired against the Meander adversary. Losses were lower with more episodes only in the Beeline adversary ($F(4.29,278.7) = 7.69, p < .001, \eta^2 = .02$) but not in the Meander ($F(4.12,214.1) = 1.256, p = .29, \eta^2 = .01$); and the number of disruptions decreased only in the Beeline adversary ($F(4.93,320.45) = 10.70, p < .001, \eta^2 = .08$) and not in the Meander ($F(6,312) = 1.95, p = .07, \eta^2 = .02$). 


The analyses above demonstrate significant differences in defense outcomes when defenders confront Beeline or Meander adversaries. The results suggest that Beeline is initially a significantly more damaging attack strategy than Meander. This makes sense by the definition of the strategy, where the Beeline adversary advances directly through the subnets to the operational sever. However, importantly, participants were able to learn the behavior of the Beeline adversary and improve their defense in a way that the loss and number of disruptions improved with more episodes in the task.  Participants were more successful against the Meander strategy; however, they were unable to significantly improve their performance with more episodes.

In what follows, we further analyze the process by which participants behaved over the course of the episodes. We analyze the participants proportion of actions, the dynamics of defense actions over time, and characterize their defense strategies. We also explore the individual differences of these behaviors.

\subsection{Process Metrics}
\subsubsection{Defense Actions}
We analyzed the defense actions taken by the participants while executing the task. Table~\ref{tab:descriptive_stat_cmd} presents the overall average proportion of use of each of the four defense actions, \textit{Analyze, Monitor, Remove, Restore}- in each of the two adversary strategies. 

     \begin{table}[!t]
        \caption{Descriptive statistics (mean $\pm$ standard deviation) regarding the average proportion of command usage per attacker type.}\label{tab:descriptive_stat_cmd}
        \begin{tabular*}{\columnwidth}{@{\extracolsep\fill}lllll@{\extracolsep\fill}}
        \toprule
         & Beeline & Meander\\
        \midrule
        Analyze  & .20 $\pm$ .14 & .19 $\pm$ .11 \\
        Monitor & .36 $\pm$ .20 & .30 $\pm$ .19 \\
        Remove & .32 $\pm$ .19 & .39 $\pm$ .22 \\ 
        Restore & .19 $\pm$ .09 & .19 $\pm$ .09 \\ 
        \botrule
        \end{tabular*}
    \end{table}

In general, the Monitor and Remove actions seem to be more popular compared to the Analyze and Restore actions among defenders, regardless of the strategy. 
ANOVAs performed for each adversary group revealed significant differences on the proportion of use of these actions when facing Beeline ($F(3,264) = 17.91, p < .001, \eta^2 = .17$) and when facing Meander ($F(3,208) = 18.80, p < .001, \eta^2 = .21$).
Post-hoc comparisons using Tukey’s HSD corrections confirm that, regardless of the type of adversary, the proportion of use of Monitor and Analyze; Monitor and Restore; Remove and Analyze; and Remove and Restore were significantly different at $p<0.001$.


Overall, participants in both conditions used Monitor and Remove actions significantly more often than Analyze and Restore\footnote{We noted a weak but significant positive correlation between the proportion of Analyze command used and the Cybersecurity background of participants (Spearman rank correlation: $r_s = .23, p = 0.011$). This correlation could explain another weak but significant negative correlation found between the participant's background and the \textit{Loss} (Spearman rank correlation: $r_s = -.27, p = .0024$). "Expert" subjects seemed to be overly focused on the Analyze action. However, the discussion of this result is beyond the scope of this paper}.

To observe the dynamics of the use of these defense actions over the course of episodes, we analyzed the proportions of actions on two levels: (1) across episodes, to observe potential learning and progressive establishment of a defense strategy, and (2) within episodes, aggregating all episodes and analyzing across the 25 steps of episodes.

\begin{figure}[!t]
    \centering
    \includegraphics[scale=.9]{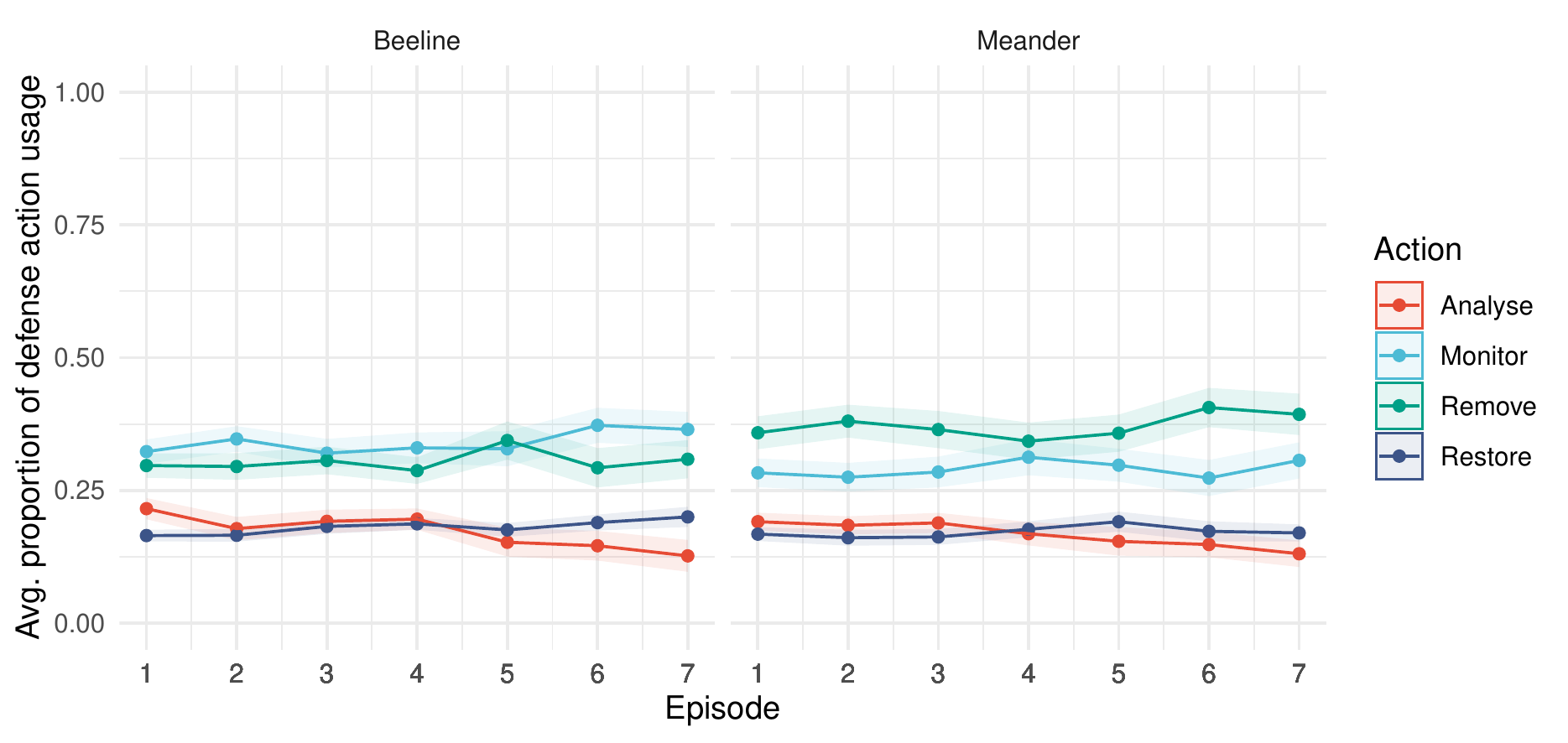}
    \caption{Average proportion of defense action usage over episodes with standard error of the mean.}
    \label{fig:FqcyEp}
\end{figure}

Fig. \ref{fig:FqcyEp} shows the average proportion of actions over the course of the seven episodes. The defender's behavior appears to be very similar in both adversary strategies across episodes. The main differences observed are that the actions Monitor and Remove are more common than the actions Analyze and Restore. In addition, the action Remove is more common when the defender confronts the Meander than when confronting the Beeline adversary.

However, mixed-effect ANOVAs on the proportion of each of the action types only revealed a significant effect of the episode on the proportion of Analyze action ($F(4.33,506.54) = 8.318,p < .001,\eta^2 = .02$) when playing against the Beeline and also the Meander adversaries. No effects of the type of adversary were found for any of the actions. 

    \begin{figure}[!t]
        \centering
        \includegraphics[scale=.9]{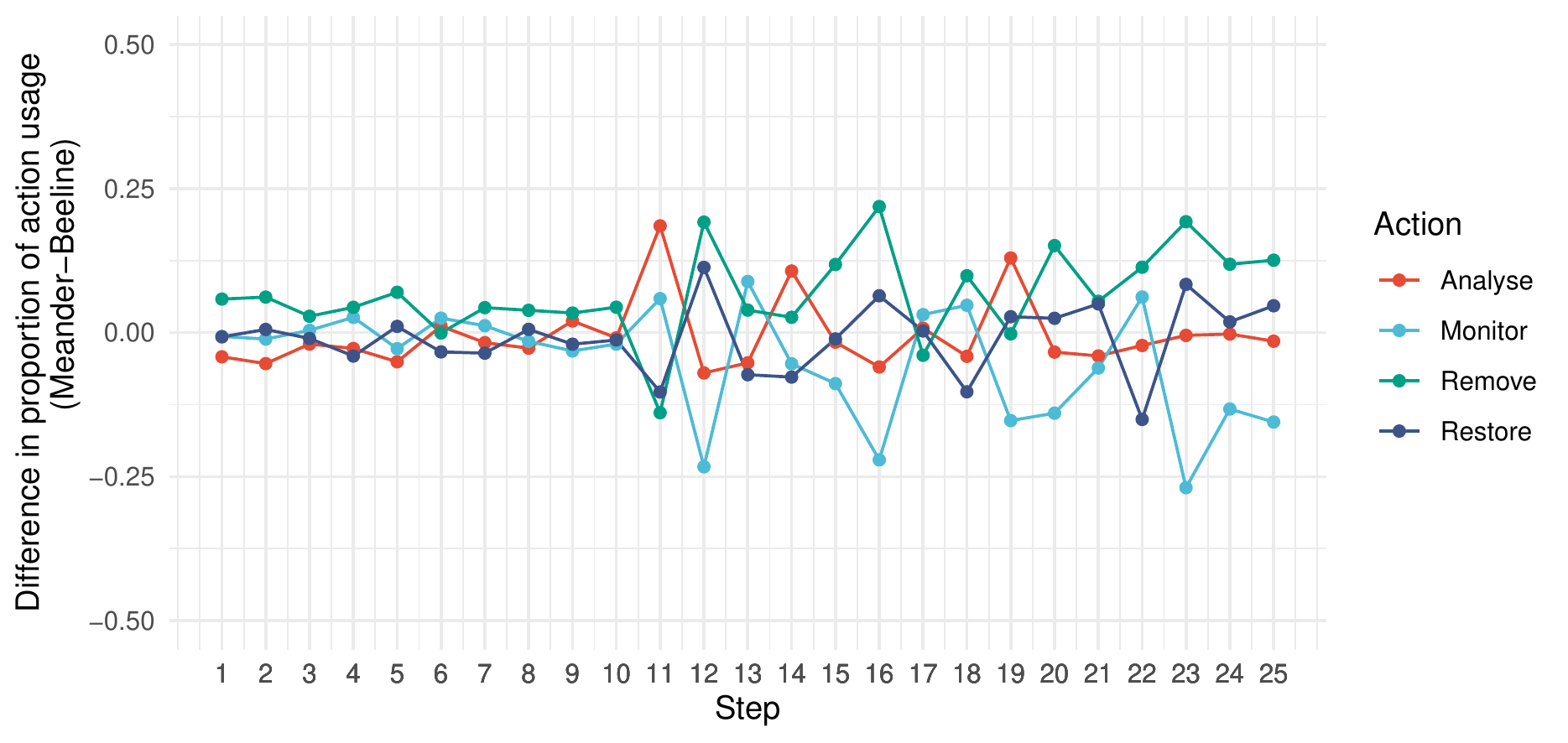}
        \caption{Difference in average proportion of action usage between Meander and Beeline conditions. A positive value indicates a higher proportion of the command in the Meander condition, a negative one indicates a higher proportion in the Beeline condition.}
        \label{fig:Diff}
    \end{figure}
    
We also analyzed the proportion of actions performed at each step over all episodes. To highlight the differences between the two adversaries, we calculated the difference between the proportion of actions taken by participants facing the Meander opponent and the proportion of actions taken by participants facing the Beeline opponent. Fig. \ref{fig:Diff} presents this difference.

We observe a larger number of Remove actions initially in the Meander compared to the Beeline, and the larger number of Analyse actions in the Beeline compared to Meander in the fist 10 steps. The difference in the proportion of actions is relatively consistent and stable during the first 10 steps. However, after step 10 we observe significant variability in this difference of the proportion of actions, noticing that the participants against the Beeline adversary engage in more Monitor actions than those playing against the Meander.

The proportion of actions against Beeline and Meander was tested for each type of action during steps 1 to 10, and then during steps 11-25. Table \ref{tab:anova_diff} indicates that the only significant difference is in the proportion of Monitor and Remove actions during steps 11-25. The proportion of Monitor actions for participants that confronted the Beeline strategy was higher than those that confronted the Meander strategy. Also, the proportion of Remove actions for participants that confronted the Meander strategy was higher than those that confronted the Beeline strategy.

    \begin{table}[!t]
        \caption{Results of the ANOVA regarding the effect of adversary type in groups of steps 1-10 and 11-25}\label{tab:anova_diff}
        \begin{tabular*}{\columnwidth}{@{\extracolsep\fill}lllllllll@{\extracolsep\fill}}
        \toprule
          & Command &  NumDF & DenDF & F value & p & p.signif & $\eta^2$\\
        \midrule
        1-10  & & & & & & &\\
          & Analyze & 1.00 & 686.40 & 3.53 & .06 &  & .08\\
          & Monitor & 1.00 & 670.47 & 0.08 & .784 & & .03\\
          & Remove & 1.00 & 610.51 & 2.61 & .107 & & .07\\
          & Restore & 1.00 & 685.28 & 0.27 & .601 & & .04\\
        11-25  & & & & & & &\\
          & Analyze & 1.00 & 1014.13 & 0.08 & .78 & & .03\\
          & Monitor & 1.00 & 1016.06 & 38.80 & $<.001$ & *** & .23\\
          & Remove & 1.00 & 992.60 & 24.47 & $<.001$ & *** & .20\\
          & Restore & 1.00 & 1025.17 & 1.72 & .191 & & .05\\
        \botrule
        *** p $< 0.001$.
        \end{tabular*}
    \end{table}

\begin{figure}[!t]
    \centering
    \includegraphics[scale=.9]{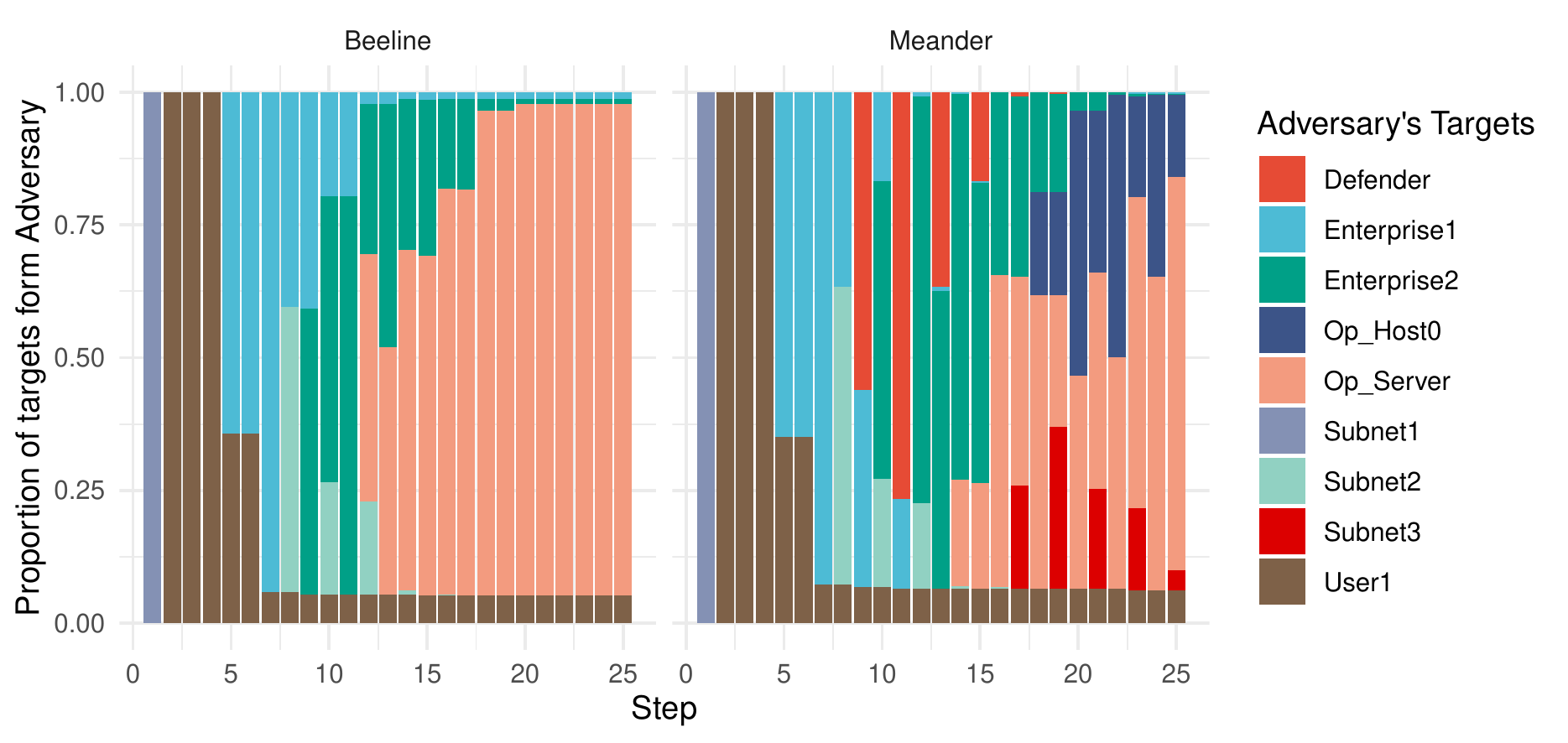}
    \caption{Evolution of the proportion of attack by target across steps.}
    \label{fig:RedTarget}
\end{figure}

To explain these defense behaviors within episodes, we analyzed the types of targets that each of the adversarial strategies attacked in each of the steps aggregated across all episodes. Fig. \ref{fig:RedTarget} represents the proportion of targets that each of the adversaries attacked on each step.

We observe that both adversaries start by attacking Subnet 1, then move to User 1, then to Enterprise 1, and then to Subnet 2. This similarity of adversarial actions appears during the first 8 steps of the game. After this step Meander starts to target different hosts, such as 'Defender', while Beeline moves on to Enterprise 2 and then directly to the Operational Server. This illustration explains the differences in the two attack strategies and explains why the human defenders' actions vary after Step 10 and differs in the Monitoring and Removing actions during steps 11-25.
  
\subsubsection{Defense Strategies}\label{DefenseStratSection}
To understand the actions taken by the defenders at a more strategic level, we classified the defense actions into three groups of strategies: \textit{Reactive}, \textit{Proactive} and \textit{Passive} strategies.
In the cyber literature, \textit{proactive} and \textit{reactive} strategies usually refer to the general approach institutions have for their cybersecurity, i.e., anticipating future threats versus patching security flaws that could expose them to known threats \citep{GrishamProact,bhuyan2020transforming,samtani2020cybersecurity,ZARREH20181255}. Here, as we focus on the operational level rather than the organizational one, we categorized each individual decision and action according to the following definition:

\begin{itemize}
    \item The \textit{reactive} strategy represents actions that result in an improved state of the network, such as the recovery of infected hosts. These are actions that the defender takes after hosts have already been attacked by the adversary and defense points have been lost. 
    
    \item The \textit{proactive} strategy is characterized by preventive actions. These are actions that reflect an anticipation of the next adversarial move or a prediction of the intention of the adversary, in a way that the defender is able to block the progression of the attack. 
    
    \item The \textit{passive} strategy represents defense actions that have no direct effect on the state of the network or slowing or stopping the progress of the adversary in the network.  
\end{itemize}

Table \ref{tab:strat_heuristics} presents the set of high-level heuristics used to categorize defense actions into one of the three strategies. 
Using the defender action, the state of the network (e.g. is the defender targeting a host that is or has been attacked), and the effect of the defense action, we coded each of these heuristics. Using this coding scheme, 91\% of all defender's actions were categorized.
    
    \begin{table}[!t]
        \caption{Heuristics.}\label{tab:strat_heuristics}
        \begin{tabular*}{\columnwidth}{@{\extracolsep\fill}ll@{\extracolsep\fill}}
        \toprule
        Behavior & Strategy\\
        \midrule
        Recovering a compromised host at the user or administrator level & Reactive \\
        Recovering the Operational Server when it is impacted & Reactive \\
        Blocking an initial Impact attempt & Proactive \\
        Preventing a host from being compromised & Proactive \\
        Repeating a successful action & Proactive \\
        Monitoring or Analyzing & Passive \\
        \botrule
        \end{tabular*}
    \end{table}

The overall proportion of reactive, proactive, and passive strategies coded from the defenders' actions when confronted with Beeline and Meander adversaries are presented in Table \ref{tab:descriptive_stat_str}. The table indicates that passive strategies are more common than proactive strategies.  

    \begin{table}[!t]
        \caption{Descriptive statistics (mean $\pm$ standard deviation) regarding the average proportion of defense strategy per attacker type.}\label{tab:descriptive_stat_str}
        \begin{tabular*}{\columnwidth}{@{\extracolsep\fill}lll@{\extracolsep\fill}}
        \toprule
         & Beeline & Meander\\
        \midrule
        Reactive  & .27 $\pm$ .15 & .26 $\pm$ .16 \\
        Proactive & .19 $\pm$ .19 & .15 $\pm$ .20 \\
        Passive & .48 $\pm$ .22 & .45 $\pm$ .24 \\ 
        \botrule
        \end{tabular*}
    \end{table}

\begin{figure}[!t]
    	\centering
    		\includegraphics[scale=.9]{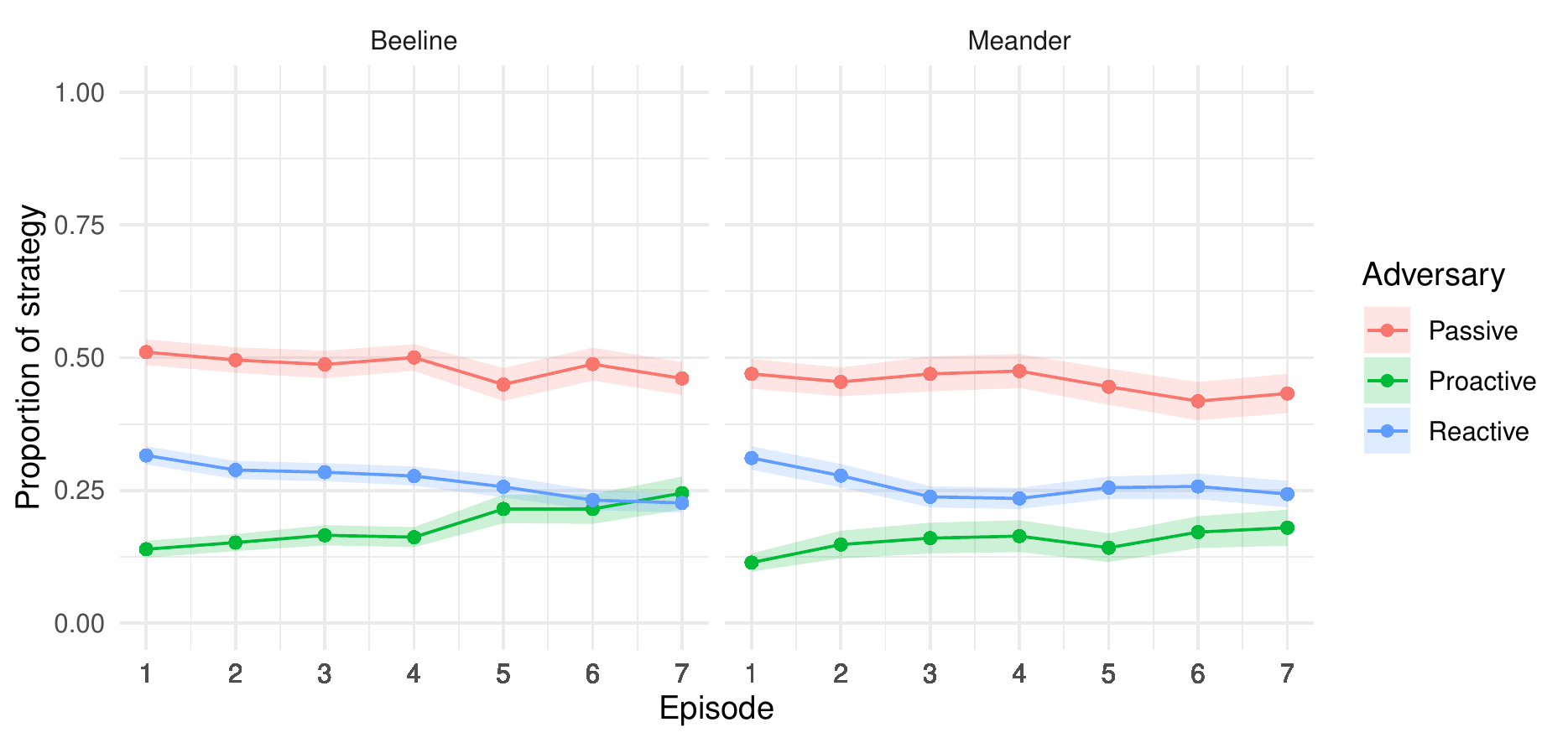}
    	\caption{Average proportion of each strategy per episode.}
    	\label{fig:Strat_ep}
    \end{figure}

Fig. \ref{fig:Strat_ep} presents the proportion of these strategies per episode. This figure illustrates that passive strategies are most common, regardless of the type of adversary. The proportion of reactive strategies decreases over the course of episodes, while the proportion of proactive strategies increases. This pattern appears to be very similar for both adversaries, although the increase of proactive strategies appears to be faster against the Beeline adversary compared to the Meander adversary.

The mixed-ANOVA results shown in Table \ref{tab:anova_MixedStrat} indicates a significant effect of the episode on the proportion of each strategy in both types of Adversaries. It also shows a significant interaction between the episode and the type of adversary for the proportion of \textit{proactive} strategy. 

\begin{table}[!t]
        \caption{Results of the mixed ANOVA regarding the effect of adversary type and episodes on the proportion of defense strategies}\label{tab:anova_MixedStrat}
        \begin{tabular*}{\columnwidth}{@{\extracolsep\fill}lllllllll@{\extracolsep\fill}}
        \toprule
        Strategy & &  NumDF & DenDF & F value & p & p.signif & $\eta^2$\\
        \midrule
        Reactive  & & & & & & &\\
          & Adversary &  1 & 117.00  & 0.18  & .675  &  & .00\\
          & Episode & 4.15 & 485.82 & 8.83 & $<.001$ & *** & .03\\
          & Adversary:Episode & 4.15 & 485.82 & 2.30  & .550 &  & .01\\
        Proactive  & & & & & & &\\
          & Adversary &  1 & 117.00  & 1.09  & .299  &  & .01\\
          & Episode & 3.03 & 354.99 & 9.23 & $<.001$ & *** & .02\\
          & Adversary:Episode & 3.03 & 354.99 & 2.70 & .045 & * & .01 \\
        Passive  & & & & & & &\\
          & Adversary & 1  & 117.00  & 0.66  &  .417 &  & .00\\
          & Episode & 3.73 & 436.85 & 3.51 & .009 & ** & .01\\
          & Adversary:Episode & 3.73 & 436.85 & 1.11 & .352 & & .00\\
        \botrule
        * p $< 0.05$, ** p $< 0.01$, *** p $< 0.001$.
        \end{tabular*}
    \end{table}

    \begin{figure}[!t]
    	\centering
    		\includegraphics[scale=.7]{fig/Fig8.pdf}
    	\caption{Proportion of each strategy per subject and episode. Subjects are ordered by Loss. Least performing subject (maximum loss) in the top left corner. The loss value is displayed above each graph.}
    	\label{fig:Strat_perEp}
    \end{figure}
    
Post-hoc 1-way ANOVAs, and considering the Bonferroni adjusted p-value (p.adj), it can be seen that the simple main effect of Episode on the proportion of Proactive strategy was significant against Beeline ($F(2.46,159.66) = 9.152,p.adj < .001,\eta^2 = .04$) but not against Meander ($F(3.11,161.83) = 2.930, p.adj = .068, \eta^2=.01$). 

\subsubsection{Individual Differences}

Fig. \ref{fig:Strat_perEp} represents the proportion of each strategy fit per episode for each individual participant separately. Furthermore, these panels are organized according the overall loss of each of the participants, where the top-left panel represents the participant with the maximum loss and the bottom-right panel represents the participant with the minimum loss. 

This figure immediately reveals the variability in the individual behaviors and the connections between the strategy that each participant used and the individual loss. Many unsuccessful defenders use passive strategies more often, while more successful defenders were more proactive.

\subsubsection{Strategy and Loss Correlations}
\begin{figure}[!h]
    	\centering
    		\includegraphics[scale=.9]{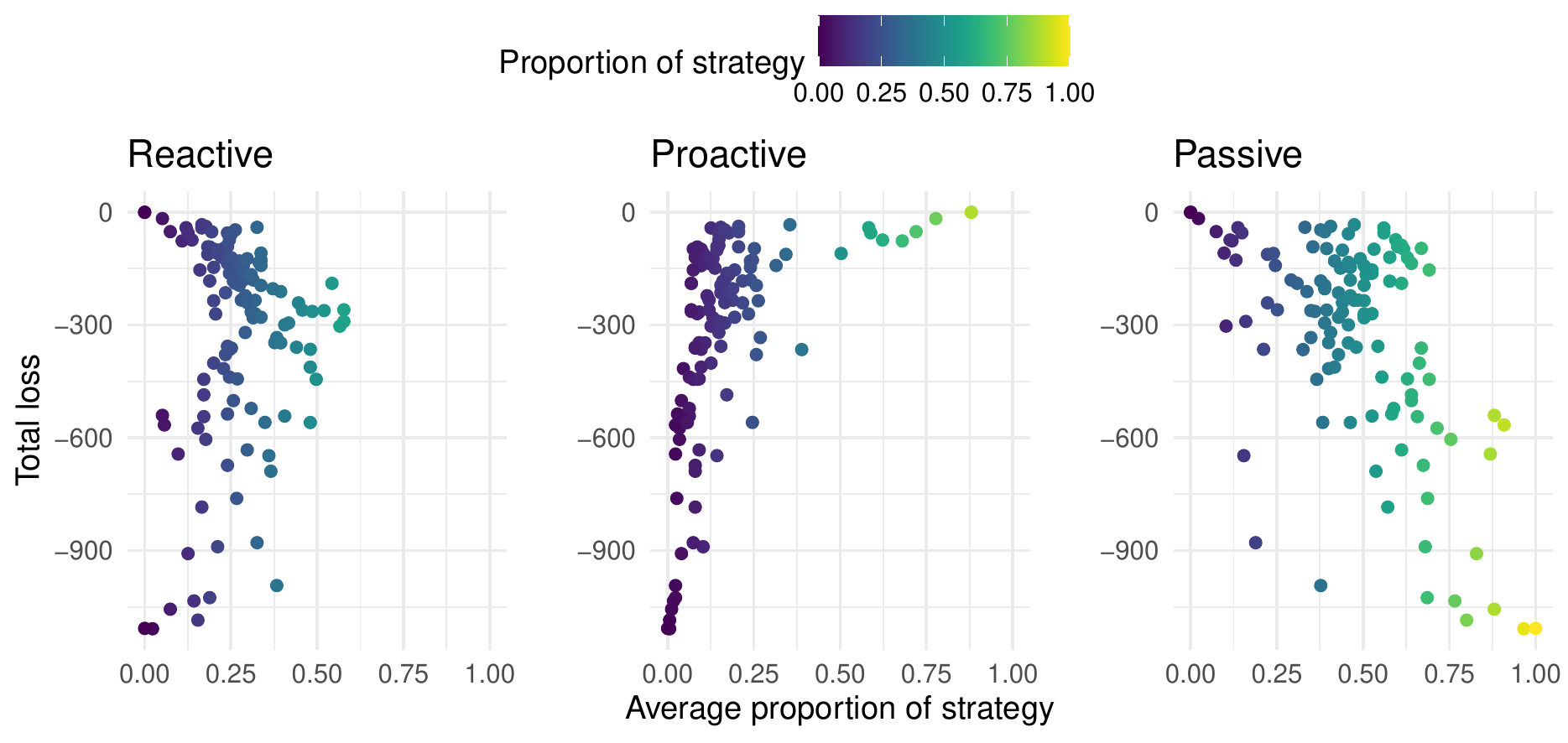}
    	\caption{Scatter plot of subject's total Loss and proportion of strategy.}
    	\label{fig:Strat_loss}
    \end{figure}

The association between the strategy and the total loss across both adversaries, was also analyzed through correlations. Scatter plots in Fig. \ref{fig:Strat_loss} represent the relationship between each individual defender's total loss score and the proportion of each strategy.

Spearman's correlation tests indicate a strong significant positive correlation between the participant's loss and the proportion of proactive strategy (Spearman rank correlation: $r_s = 0.66, p < .001$). That is, generally, defenders with a higher proportion of proactive behaviors are more likely to lose fewer points, i.e., to protect the network better. Being proactive, such as performing a Remove action that prevents a host from being exploited, is an efficient way to prevent loses and being more successful in protecting the network.

Similarly, Spearman's correlation tests indicate a moderate significant negative correlation between the defender's loss and its proportion of passive strategy (Spearman rank correlation: $r_s = -0.45, p < .001$). Defenders with larger number of passive actions were more likely to lose more points since they are not taking any active defense action, i.e., they are not protecting the network.  

Finally, the correlation between the defender's loss and the proportion of reactive strategy was not significant.


\section{Discussion}

We designed a simple cyber defense game as a web-based application, to study human defense decisions against simulated adversaries. In this experiment, we measured the impact of two different deterministic attack strategies on defenders' behaviors. To do so, we analyzed their performance, their defense choices and behaviors, and their strategies.


As expected, the defenders performance reflects the difference in "aggressiveness” of the attack strategy in terms of Loss, Recovery Time and number of Disruptions.
Indeed, as an attacker following the Beeline strategy was quicker to reach the Operational Server than one following a Meander strategy, it resulted in significantly bigger Loss for the human defender, more Disruptions and longer Recovery Time.
However, we have observed that, over the episodes and independently from the condition, participants have managed to improve their performance and lower their Loss. Two possible explanations can be investigated for the overall improvement: (1) the number of Disruptions dropped while subjects learned to more efficiently prevent the attacker from reaching the Operational Server and/or, (2) the Recovery Time improved, i.e. subjects became faster to recover the Operational server from a disruption.

Results indicate a significant drop in the number of Disruptions recorded over time, while no amelioration is noticeable in terms of Recovery Time.
This can be interpreted as the defenders learning to more efficiently block the progression of the attacker in the network, before it reaches the Operational Server.

Overall, participants confronted with a Beeline attacker learned to develop an efficient Proactive defense strategy to improve their performance, be it in terms of loss, number of disruptions and recovery time. Our interpretation is that, even though both attack strategies are deterministic, Beeline is more direct and consistent, and routing through a smaller number of hosts than Meander. This makes the Beeline strategy easier for the defenders to build a mental representation and and to predict the adversarial actions with increased defense experience. The predictability of the strategy of attack had a significant influence on how humans learn an effective defense strategy.  

Although participants that faced the Beeline adversary seemed to significantly improve their performance over time, they only succeeded to achieve similar level of performance than participants that faced the Meander adversary.
In some ways, the Beeline adversary leaves more room for improvement, which could also be a factor in the observed difference in learning pace. In past results involving experiments with cognitive models on the same task \citep{hfesIBL}, defense agents showed accentuated learning curves when confronted to a Beeline attacker but similar final performance after a large number of episodes. It would be interesting to see how humans are able to improve their strategies and how their performance evolves with more episodes.
Also, in future work, longer episodes (i.e., more than 25 steps) could allow us to use patterns identification methods and extended analysis of actions sequences, to refine the categorization of defense strategies and perhaps identify more complex heuristics.


In general, this study illustrates how the type of simulated adversary that human defenders face may influence the speed of learning and the development of an adequate defense strategy. A more aggressive but more predictive attacker was found to be easier to learn and exploit by human defender compared to a stealthy and less predictable adversary.

Cyber analysts have to work in a highly dynamic environment, with flawed and noisy information. Adversarial cyber defense games and simulation tools like the IDG can help simulate such decision-making situations and better understand the cognitive demands faced by humans cyber defenders. Our work shed light on the importance of providing dynamic and complex attackers for the development and training of human defenders. These results support the findings of recent modeling experiments that have shown that dynamic attack strategies are a weakness for cognitive models and AI defense \citep{hfesIBL,hicssIBL}.
To progress towards building human and AI collaboration in cyber defense future work, we may look into the effect of dynamic attackers on the human development of defense strategies, and investigate how humans can work within teams of AI agents and collaborate with them.

\section{Competing interests}
No competing interest is declared.

\section{Author contributions statement}
B.P.: Conceptualization of this study, Methodology, Software development, Data analysis, Writing; Y.D.: Conceptualization of the study, Software development, Writing C.G.: Conceptualization of the study, Methodology, Writing - Original draft preparation.

\section{Acknowledgments}
The authors thank the anonymous reviewers for their valuable suggestions. We thank Jeffrey Flagg, Dynamic Decision Making Laboratory, for research assistance in reviewing and running the study. This research was sponsored by the Army Research Office and accomplished under Australia-US MURI Grant Number W911NF-20-S-000 and by the Army Research Laboratory under Cooperative Agreement Number W911NF-13-2-0045 (ARL Cyber Security CRA).

\bibliographystyle{unsrt}

\renewcommand{\bibnumfmt}[1]{[#1] }

\bibliography{oup-authoring-template}

\begin{thebibliography}{10}

\bibitem{li2021comprehensive}
Yuchong Li and Qinghui Liu.
\newblock A comprehensive review study of cyber-attacks and cyber security;
  emerging trends and recent developments.
\newblock {\em Energy Reports}, 7:8176--8186, 2021.

\bibitem{thanh2019survey}
Cong~Truong Thanh and Ivan Zelinka.
\newblock A survey on artificial intelligence in malware as next-generation
  threats.
\newblock In {\em Mendel}, volume~25, pages 27--34, 2019.

\bibitem{colbert2020game}
Edward~JM Colbert, Alexander Kott, and Lawrence~P Knachel.
\newblock The game-theoretic model and experimental investigation of cyber
  wargaming.
\newblock {\em The Journal of Defense Modeling and Simulation}, 17(1):21--38,
  2020.

\bibitem{ferguson2018tularosa}
Kimberly Ferguson-Walter, Temmie Shade, Andrew Rogers, Michael
  Christopher~Stefan Trumbo, Kevin~S Nauer, Kristin~Marie Divis, Aaron Jones,
  Angela Combs, and Robert~G Abbott.
\newblock The tularosa study: An experimental design and implementation to
  quantify the effectiveness of cyber deception.
\newblock Technical report, Sandia National Lab.(SNL-NM), Albuquerque, NM
  (United States), 2018.

\bibitem{applebaum2016intelligent}
Andy Applebaum, Doug Miller, Blake Strom, Chris Korban, and Ross Wolf.
\newblock Intelligent, automated red team emulation.
\newblock In {\em Proceedings of the 32nd Annual Conference on Computer
  Security Applications}, pages 363--373, 2016.

\bibitem{kavak2021simulation}
Hamdi Kavak, Jose~J Padilla, Daniele Vernon-Bido, Saikou~Y Diallo, Ross Gore,
  and Sachin Shetty.
\newblock Simulation for cybersecurity: state of the art and future directions.
\newblock {\em Journal of Cybersecurity}, 7(1):tyab005, 2021.

\bibitem{varshney2011live}
Maneesh Varshney, Kent Pickett, and Rajive Bagrodia.
\newblock A live-virtual-constructive (lvc) framework for cyber operations
  test, evaluation and training.
\newblock In {\em 2011-MILCOM 2011 Military Communications Conference}, pages
  1387--1392. IEEE, 2011.

\bibitem{gutzwiller2016task}
Robert~S Gutzwiller, Sarah~M Hunt, and Douglas~S Lange.
\newblock A task analysis toward characterizing cyber-cognitive situation
  awareness (ccsa) in cyber defense analysts.
\newblock In {\em 2016 IEEE International Multi-Disciplinary Conference on
  Cognitive Methods in Situation Awareness and Decision Support (CogSIMA)},
  pages 14--20. IEEE, 2016.

\bibitem{veksler2020cognitive}
Vladislav~D Veksler, Norbou Buchler, Claire~G LaFleur, Michael~S Yu, Christian
  Lebiere, and Cleotilde Gonzalez.
\newblock Cognitive models in cybersecurity: learning from expert analysts and
  predicting attacker behavior.
\newblock {\em Frontiers in Psychology}, 11:1049, 2020.

\bibitem{veksler2018simulations}
Vladislav~D Veksler, Norbou Buchler, Blaine~E Hoffman, Daniel~N Cassenti, Char
  Sample, and Shridat Sugrim.
\newblock Simulations in cyber-security: a review of cognitive modeling of
  network attackers, defenders, and users.
\newblock {\em Frontiers in psychology}, 9:691, 2018.

\bibitem{cranford2021towards}
Edward~A Cranford, Cleotilde Gonzalez, Palvi Aggarwal, Milind Tambe, Sarah
  Cooney, and Christian Lebiere.
\newblock Towards a cognitive theory of cyber deception.
\newblock {\em Cognitive Science}, 45(7):e13013, 2021.

\bibitem{johnson2021decision}
Chelsea~K Johnson, Robert~S Gutzwiller, Joseph Gervais, and Kimberly~J
  Ferguson-Walter.
\newblock Decision-making biases and cyber attackers.
\newblock In {\em 2021 36th IEEE/ACM International Conference on Automated
  Software Engineering Workshops (ASEW)}, pages 140--144. IEEE, 2021.

\bibitem{gonzalez2014cognition}
Cleotilde Gonzalez, Noam Ben-Asher, Alessandro Oltramari, and Christian
  Lebiere.
\newblock Cognition and technology.
\newblock In {\em Cyber defense and situational awareness}, pages 93--117.
  Springer, 2014.

\bibitem{Triad_JONES2021106799}
Daniel~N. Jones, Edgar Padilla, Shelby~R. Curtis, and Christopher Kiekintveld.
\newblock Network discovery and scanning strategies and the dark triad.
\newblock {\em Computers in Human Behavior}, 122:106799, 2021.

\bibitem{Triad_CURTIS2018174}
Shelby~R. Curtis, Prashanth Rajivan, Daniel~N. Jones, and Cleotilde Gonzalez.
\newblock Phishing attempts among the dark triad: Patterns of attack and
  vulnerability.
\newblock {\em Computers in Human Behavior}, 87:174--182, 2018.

\bibitem{gutzwiller2015human}
Robert~S Gutzwiller, Sunny Fugate, Benjamin~D Sawyer, and PA~Hancock.
\newblock The human factors of cyber network defense.
\newblock In {\em Proceedings of the human factors and ergonomics society
  annual meeting}, volume~59, pages 322--326. SAGE publications Sage CA: Los
  Angeles, CA, 2015.

\bibitem{BUCHLER2018}
Norbou Buchler, Prashanth Rajivan, Laura~R. Marusich, Lewis Lightner, and
  Cleotilde Gonzalez.
\newblock Sociometrics and observational assessment of teaming and leadership
  in a cyber security defense competition.
\newblock {\em Computers \& Security}, 73:114--136, 2018.

\bibitem{strom2018mitre}
Blake~E Strom, Andy Applebaum, Doug~P Miller, Kathryn~C Nickels, Adam~G
  Pennington, and Cody~B Thomas.
\newblock Mitre attack: Design and philosophy.
\newblock In {\em Technical report}. The MITRE Corporation, 2018.

\bibitem{gonzalez2005use}
Cleotilde Gonzalez, Polina Vanyukov, and Michael~K Martin.
\newblock The use of microworlds to study dynamic decision making.
\newblock {\em Computers in human behavior}, 21(2):273--286, 2005.

\bibitem{aggarwal2020hackit}
Palvi Aggarwal, Cleotilde Gonzalez, and Varun Dutt.
\newblock Hackit: a real-time simulation tool for studying real-world
  cyberattacks in the laboratory.
\newblock In {\em Handbook of Computer Networks and Cyber Security}, pages
  949--959. Springer, 2020.

\bibitem{singh2019training}
Kuldeep Singh, Palvi Aggarwal, Prashanth Rajivan, and Cleotilde Gonzalez.
\newblock Training to detect phishing emails: Effects of the frequency of
  experienced phishing emails.
\newblock In {\em Proceedings of the human factors and ergonomics society
  annual meeting}, volume~63, pages 453--457. SAGE Publications Sage CA: Los
  Angeles, CA, 2019.

\bibitem{ben2015effects}
Noam Ben-Asher and Cleotilde Gonzalez.
\newblock Effects of cyber security knowledge on attack detection.
\newblock {\em Computers in Human Behavior}, 48:51--61, 2015.

\bibitem{moisan2017security}
Fr{\'e}d{\'e}ric Moisan and Cleotilde Gonzalez.
\newblock Security under uncertainty: adaptive attackers are more challenging
  to human defenders than random attackers.
\newblock {\em Frontiers in psychology}, 8:982, 2017.

\bibitem{hutchins2011intelligence}
Eric~M Hutchins, Michael~J Cloppert, Rohan~M Amin, et~al.
\newblock Intelligence-driven computer network defense informed by analysis of
  adversary campaigns and intrusion kill chains.
\newblock {\em Leading Issues in Information Warfare \& Security Research},
  1(1):80, 2011.

\bibitem{zhang2021three}
Li~Zhang and Vrizlynn~LL Thing.
\newblock Three decades of deception techniques in active cyber
  defense-retrospect and outlook.
\newblock {\em Computers \& Security}, 106:102288, 2021.

\bibitem{tambe2011security}
Milind Tambe.
\newblock {\em Security and game theory: algorithms, deployed systems, lessons
  learned}.
\newblock Cambridge university press, 2011.

\bibitem{abbasi2016know}
Yasaman Abbasi, Debarun Kar, Nicole~D Sintov, Milind Tambe, Noam Ben-Asher, Don
  Morrison, and Cleotilde Gonzalez.
\newblock Know your adversary: Insights for a better adversarial behavioral
  model.
\newblock In {\em CogSci}, 2016.

\bibitem{aggarwal2015cyber}
Palvi Aggarwal, Zahid Maqbool, Antra Grover, VS~Chandrasekhar Pammi, Saumya
  Singh, and Varun Dutt.
\newblock Cyber security: A game-theoretic analysis of defender and attacker
  strategies in defacing-website games.
\newblock In {\em 2015 International Conference on Cyber Situational Awareness,
  Data Analytics and Assessment (CyberSA)}, pages 1--8. IEEE, 2015.

\bibitem{nochenson2012simulation}
Alan Nochenson and CF~Heimann.
\newblock Simulation and game-theoretic analysis of an attacker-defender game.
\newblock In {\em International Conference on Decision and Game Theory for
  Security}, pages 138--151. Springer, 2012.

\bibitem{do2017game}
Cuong~T Do, Nguyen~H Tran, Choongseon Hong, Charles~A Kamhoua, Kevin~A Kwiat,
  Erik Blasch, Shaolei Ren, Niki Pissinou, and Sundaraja~Sitharama Iyengar.
\newblock Game theory for cyber security and privacy.
\newblock {\em ACM Computing Surveys (CSUR)}, 50(2):1--37, 2017.

\bibitem{attiah2018game}
Afraa Attiah, Mainak Chatterjee, and Cliff~C Zou.
\newblock A game theoretic approach to model cyber attack and defense
  strategies.
\newblock In {\em 2018 IEEE International Conference on Communications (ICC)},
  pages 1--7. IEEE, 2018.

\bibitem{wang2016survey}
Yuan Wang, Yongjun Wang, Jing Liu, Zhijian Huang, and Peidai Xie.
\newblock A survey of game theoretic methods for cyber security.
\newblock In {\em 2016 IEEE First International Conference on Data Science in
  Cyberspace (DSC)}, pages 631--636. IEEE, 2016.

\bibitem{hfesIBL}
Yinuo Du, Baptiste Prébot, Xiaoli Xi, and Cleotilde Gonzalez.
\newblock Towards autonomous cyber defense: Predictions from a cognitive model.
\newblock {\em Proceedings of the Human Factors and Ergonomics Society Annual
  Meeting}, 2022.

\bibitem{Gonzlez2003InstancebasedLI}
Cleotilde Gonzalez, Francis~J. Lerch, and Christian Lebiere.
\newblock Instance-based learning in dynamic decision making.
\newblock {\em Cogn. Sci.}, 27:591--635, 2003.

\bibitem{GrishamProact}
John Grisham, Sagar Samtani, Mark Patton, and Hsinchun Chen.
\newblock Identifying mobile malware and key threat actors in online hacker
  forums for proactive cyber threat intelligence.
\newblock In {\em 2017 IEEE International Conference on Intelligence and
  Security Informatics (ISI)}, pages 13--18, 2017.

\bibitem{bhuyan2020transforming}
Soumitra~Sudip Bhuyan, Umar~Y Kabir, Jessica~M Escareno, Kenya Ector, Sandeep
  Palakodeti, David Wyant, Sajeesh Kumar, Marian Levy, Satish Kedia, Dipankar
  Dasgupta, et~al.
\newblock Transforming healthcare cybersecurity from reactive to proactive:
  current status and future recommendations.
\newblock {\em Journal of medical systems}, 44(5):1--9, 2020.

\bibitem{samtani2020cybersecurity}
Sagar Samtani, Maggie Abate, Victor Benjamin, and Weifeng Li.
\newblock Cybersecurity as an industry: A cyber threat intelligence
  perspective.
\newblock {\em The Palgrave Handbook of International Cybercrime and
  Cyberdeviance}, pages 135--154, 2020.

\bibitem{ZARREH20181255}
Alireza Zarreh, Can Saygin, HungDa Wan, Yooneun Lee, and Alejandro Bracho.
\newblock A game theory based cybersecurity assessment model for advanced
  manufacturing systems.
\newblock {\em Procedia Manufacturing}, 26:1255--1264, 2018.
\newblock 46th SME North American Manufacturing Research Conference, NAMRC 46,
  Texas, USA.

\bibitem{hicssIBL}
Yinuo Du, Baptiste Prébot, and Cleotilde Gonzalez.
\newblock A cyber-war between bots: Human-like attackers are more challenging
  for defenders than deterministic attackers.
\newblock {\em Accepted to the 56th Hawaii International Conference on System
  Sciences HICSS 2023}, 2023.

\end{thebibliography}


\begin{biography}{}{\author{Author Name.} This is sample author biography text this is sample author biography text this is sample author biography text this is sample author biography text this is sample author biography text this is sample author biography text this is sample author biography text this is sample author biography text.}
\end{biography}

\end{document}